\documentclass[preprint,aps,showpacs,preprintnumbers,amsmath,amssymb]{revtex4}
\usepackage{graphicx}
\usepackage{dcolumn}
\usepackage{bm}

\begin{document}

\title{Resonance energy of the $\bar{K}NN-\pi Y N$ system }

\author{Y. Ikeda and T. Sato}
\email{ikeda@kern.phys.sci.osaka-u.ac.jp, tsato@phys.sci.osaka-u.ac.jp}
\affiliation{%
Department of Physics, Graduate School of Science,
Osaka University, Toyonaka, Osaka 560-0043, Japan}%

\date{\today}

\begin{abstract}
The resonance energies of strange dibaryons
are investigated with the use of the $\bar{K}NN-\pi Y N $ 
coupled-channels Faddeev equation.
It is found  that the pole positions of 
the predicted three-body amplitudes
are significantly modified 
when the three-body  coupled-channels dynamics is approximated,
as is done in the literature,
by the effective two-body $\bar{K}N$ interactions.

\end{abstract}


\maketitle

\section{Introduction}
\label{intro}
Since  the deeply bound kaonic nuclear states were
 predicted\cite{Aka1,Aka2,Dote}, the
 few-nucleon systems with strangeness 
have attracted  increasing interest.
It is generally believed that those states can be generated
by an attractive interaction between a kaon and a nucleon 
in the isospin $I=0$ channel. 
It has been suggested 
that the kaon-nucleon interaction may even modify 
the  spatial distribution of nucleons in nuclei.
Among the deeply bound kaonic states, the
resonances in  the $\bar{K}NN-\pi YN(Y=\Sigma,\Lambda)$ system
(strange dibaryon resonance) are 
particularly interesting 
since the three-hadron dynamics involved
can be handled accurately with the use of 
the well-established Faddeev equation.
This also means that
the study of  this strange dibaryon system will provide us 
 with information on the nature of the $\bar{K}N-\pi\Sigma$
 interaction
and the basic mechanism of the kaon-nucleus interactions.

 The FINUDA collaboration\cite{Agne} reported a signature of the 
$\bar{K}NN-\pi Y N$ coupled-channels resonance 
below   the  $\pi\Sigma N$ threshold. 
The reported resonance  has a binding energy  $B\sim 115$ MeV
and a width $\Gamma \sim 67$ MeV.
However 
the interpretation of the signal
is still open to discussion~\cite{Magas}.
More information on the $\bar{K}NN -\pi Y N$ resonance 
is expected to become available from
Spring-8 and J-PARC in the near future.
The first theoretical prediction of  the resonance energy was given
in Ref. \cite{Aka2}. Using a variational approach and a
phenomenological $\bar{K}N-\pi\Sigma$ potential, it was found that the 
binding energy and the width of  the $\bar{K}NN-\pi Y N$ system  are 
$(B,\Gamma) \sim (48,60)$MeV.
The calculation in Ref. \cite{DHW},
which  used a  $\bar{K}N$
interaction generated from a chiral unitary model,
gave $(B,\Gamma) \sim (20, 40 \sim 70)$MeV.
In both of these earlier works, the
three-body $\bar{K}NN-\pi Y N$ coupled-channels
problem was 
handled
with the use of effective $\bar{K}N$ interactions
obtained by truncating the Fock space into $\bar{K}NN$.
Meanwhile, 
the three-body dynamics
can be fully taken into account with the use of
the Faddeev formulation,
and a study based on the Faddeev formulation
was presented by the present authors~\cite{ikeda},
and by Shevchenko et al.~\cite{she}.
Ref.~\cite{ikeda}, which employed
the $\bar{K}N-\pi\Sigma$ interaction 
based on the leading order chiral Lagrangian,
gave $(B,\Gamma) \sim (60\sim 95, 45 \sim 80)$ MeV,
while Ref.~\cite{she}, which adopted
a phenomenological $\bar{K}N$ interaction,
reported $(B,\Gamma) \sim (50 \sim 70, 100)$MeV.
Thus, at present, theoretical predictions 
on the resonance energy
spread over a rather 
wide range~\cite{Aka2,DHW,ikeda,she,AOY,NK,WG}. 

A major uncertainty in theoretically estimating
the resonance energy of $\bar{K}NN-\pi Y N$ system 
is that an accurate description of the $\bar{K}N$ interaction
including its off-shell behavior is still missing;
this is particularly true for the the energy region below the
$\bar{K}N$ threshold.
Taking a reverse viewpoint, 
we may hope that there is a possibility
to constrain $\bar{K}N$ dynamics
from the study of the $\bar{K}NN-\pi Y N$ resonance.
To achieve this goal, however, 
it is crucial to treat
the three-body dynamics as accurately as possible
in a theoretical calculation for a given $\bar{K}N$ model.

In most of the existing theoretical work, 
the resonance energy
is predicted to lie below the $\bar{K}NN$ threshold and 
above the $\pi \Sigma N $ threshold;
thus the relevant state is a continuum (localized) 
state in the $\pi\Sigma N$ ($\bar{K}NN$) Fock space.
In this circumstance it is 
an inviting idea
to work in the $\bar{K}NN$ sub-space 
by eliminating the $\pi\Sigma N$ states
\cite{Aka2,DHW,HW}, and 
many analyses in the literature 
adopt this `effective potential approach'
and introduce effective $\bar{K}N$ interactions
to subsume the effects of the eliminated channel.
One thing to be emphasized here is that, 
when Fock space is truncated,
the resulting effective interaction in a
sub-space in general becomes a many-body operator, 
but that this fundamental feature is ignored in
the existing effective potential treatments,
which only consider  
effective two-body $\bar{K}N$ interactions.
In this connection, 
it seems worth noting that 
the Faddeev approach\cite{ikeda,she}, which fully
takes account of coupled-channels three-body dynamics, 
tends to give a deeper binding energy 
than the approximate effective potential approach.

In this report we present a detailed examination
of the nature of the approximations involved 
in the existing effective potential approach
calculations.  (For convenience, 
the approximate effective
potential approach in question will be simply 
referred to as `EPA'). 
It turns out (see below) that,
starting from the full coupled-channels 
Faddeev equations,
we can simulate `EPA'
by introducing certain simplifying assumptions 
regarding the two-body t-matrix 
embedded in the three-body system. 
This allows us to scrutinize the nature of 
approximations involved in `EPA'
in relation to the full Faddeev 
calculation~\cite{ikeda},
and to assess the validity (or limitation)
of `EPA'.
For this assessment, 
we focus here on comparison 
of the resonance positions
obtained in `EPA' and in the full calculation.

A comment is in order here concerning 
methods used to determine the resonance energy. 
In the present article we determine
the resonance energy from the position of a pole in the
scattering amplitude, 
as explained in detail in Ref.~\cite{ikeda}.
Recently, it has been suggested~\cite{Aka3} 
that a pole in the complex energy plane may 
not adequately characterize a resonance.
According to Ref.~\cite{Aka3}, 
as the strength of the $\bar{K}N$ interaction 
is artificially increased,
the trajectory of the pole
moves below the $\pi\Sigma$ threshold, but keeping 
a finite width 
(signature for a virtual state;
see however Ref. \cite{cieply}).
We shall address here this question as well
and show that  the problem of a pole 
moving to a virtual state for the three-body amplitude
is an artifact of the approximation used in `EPA'.

In section II, we briefly explain the method we use for solving
the Faddeev equation for the coupled 
$\bar{K}NN \oplus \pi Y N (Y=\Sigma, \Lambda)$ system,
and we elucidate what approximations are involved
in going from the full Faddeev formulation to `EPA'.
Section III is devoted to
the explanation of 
how the $\bar{K}N$-$\pi Y$ interactions used in our calculations
are derived from the chiral Lagrangian.
The numerical results on the predicted resonance energies
are presented in section IV,
and section V gives summary.

\section{Coupled-channel approach for $\bar{K}NN-\pi YN$ system}
\subsection{AGS equation and resonance pole}

The Faddeev equation for a  three-particle system with separable 
two-body  interactions
can be cast into 
the Alt-Grassberger-Sandhas(AGS) equation~\cite{ags},
\begin{eqnarray}
X_{i,j}(\vec{p}_i,\vec{p}_j,W) & = & 
(1-\delta_{i,j})Z_{i,j}(\vec{p}_i,\vec{p}_j,W) + 
\sum _{n\neq i} \int d \vec{p}_n 
    Z_{i,n}(\vec{p}_i,\vec{p}_n,W)
   \tau _n (W)
X_{n,j}(\vec{p}_n,\vec{p}_j,W),\nonumber \\
 \label{Eqags-2}
\end{eqnarray}
 where $W$ is the total scattering energy,
$X_{i,j}(\vec{p}_i,\vec{p}_j,W)$ with $i,j = 1,2,3$ are the 
scattering amplitudes.
The channel $i$ ($j$) is characterized 
by the spectator particle $i$ ($j$).
For example, $i=1$   represents a  quasi two-body
channel in which  the particle $1$
is the  spectator of  the last interaction between
 particles $2$ and $3$.
The momentum of the spectator particle $i$ and
the relative momentum for channel $i$ are denoted
by $\vec{p}_i$ and  $\vec{q}_i$, respectively.

\begin{figure*}
\resizebox{0.8\textwidth}{!}{%
\includegraphics{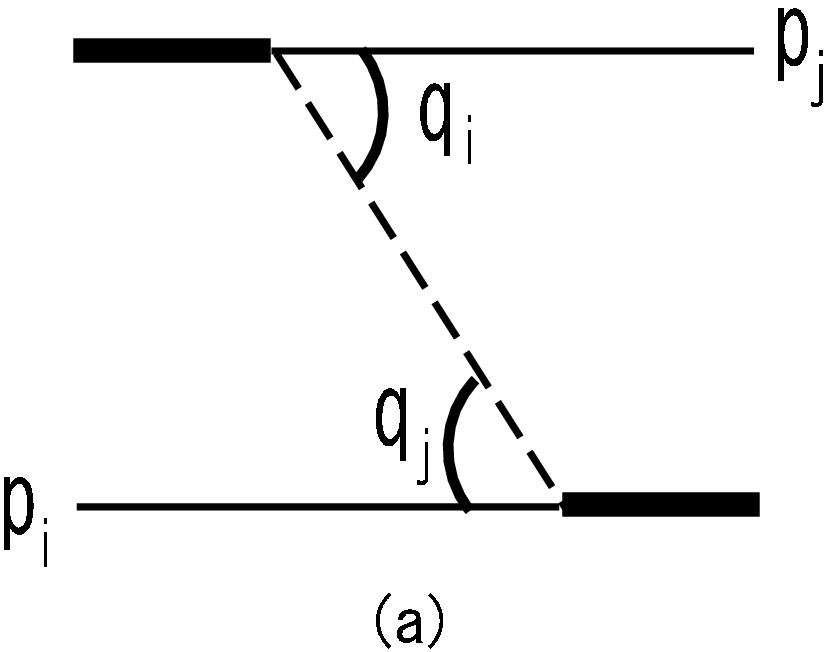}\hspace*{2cm}
\includegraphics{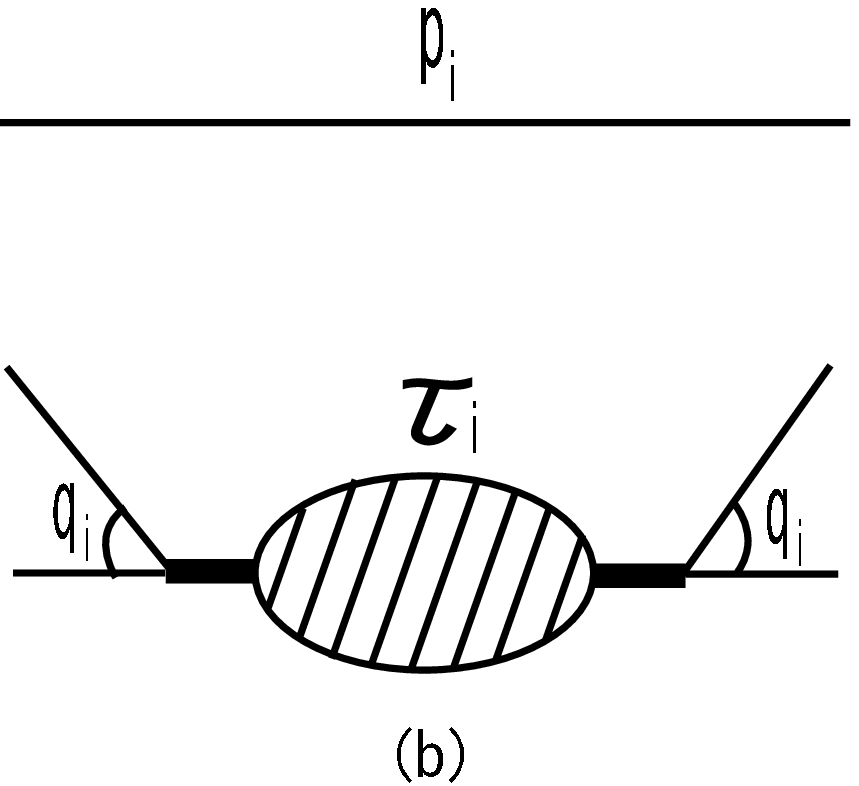}}
\caption{
Graphical representation of
(a) one-particle-exchange interaction
$Z_{i,j}(\vec p_i,\vec p_j,W)$ 
and (b) two-body t-matrix  $\tau _i(W)$.
The relative momentum of the interacting particles
is  denoted by $\vec{q}_i$ for the spectator particle $i$.
}
\label{diag-Ztau}
\end{figure*}

The driving term $Z_{i,j}(\vec{p}_i,\vec{p}_j,W) $ of the AGS equation 
is given by the  particle exchange interaction illustrated in
Fig. \ref{diag-Ztau} (a) and can be written as
\begin{eqnarray}
Z_{i,j}(\vec{p}_i,\vec{p}_j,W)  & = &
\frac{g_i^*(\vec{q}_i) g_j(\vec{q}_j)}
{W- E_i(\vec{p}_i) -E_j(\vec{p}_j) -E_{k}(-\vec{p}_i-\vec{p}_j)
 + i \epsilon}. \label{Z}
\end{eqnarray}
Here we used the two-body interaction for channel $i$ of the following form
\begin{eqnarray}
<\vec{q}_i'|v_i|\vec{q}_i> & = & \gamma_i g_i(\vec{q}_i')g_i(\vec{q}_i).
\end{eqnarray}
The `isobar' propagator $\tau_i$, illustrated in
Fig. \ref{diag-Ztau}(b), is given by
\begin{eqnarray}
(\tau_i (W))^{-1} & = & 1/\gamma_i - \int d\vec{q}_i 
\frac{  |g_i(\vec{q}_i)|^2}
{W - E_i(\vec{p}_i) - E_{jk}(\vec{p}_i,\vec{q}_i) + i\epsilon}.
\label{tau}
\end{eqnarray}
Here 
$E_{jk}(\vec{p}_i,\vec{q}_i)=
\sqrt{(E_j(\vec{q}_i)+E_k(\vec{q}_i))^2  + \vec{p}_i^{\ 2}}$
is the energy of the interacting particles ($j$ and $k$)
expressed in terms of the relative momentum $\vec{q}_i$ and 
the momentum of the spectator particle $\vec{p}_i$.
The `isobar' propagator $\tau_i$, which is a part of the two-body 
t-matrix within the three-particle system will be further examined 
in the next section.

In this work we investigate a strange dibaryon resonance with
angular momentum $J^{\pi}=0^-$ and isospin $I=1/2$.  
The main Fock-space components of the resonance are
$\bar{K}NN$ and $\pi\Sigma N$ states 
which couple with each other through the
 $I=0$ $\bar{K}N-\pi\Sigma$ interaction.  
We also take into account the $\pi\Lambda N$ component, which couples
with the main $\bar{K}N-\pi\Sigma$ components through the
$I=1$ $\bar{K}N-\pi Y$ interaction.
The AGS equation then becomes coupled-channels equations 
involving the channels: $\bar{K}NN$ 
($\bar{K}N_{I=0},\bar{K}N_{I=1},NN_{I=1}$),
$\pi \Sigma N$ ($\pi N_{I=1/2},\pi N_{I=3/2}$,
$\pi\Sigma_{I=0},\pi\Sigma_{I=1}$) and
$\pi \Lambda N$ ($\pi N_{I=1/2}, \pi\Lambda_{I=1}$).
We assume that all the orbital angular momenta 
are s-wave.
After isospin-angular momentum projection
and the anti-symmetrization of the
two nucleons, Eq. (\ref{Eqags-2}) becomes
the following coupled integral equations
\cite{book,Afnan2}
\begin{eqnarray}
X_{\alpha,\beta}(p',p,W)
 & = & C^1_{\alpha,\beta}Z_{\alpha,\beta}(p',p,W)
 + \sum_{\gamma,\delta} \int dq q^2
 C^2_{\alpha,\gamma}Z_{\alpha,\gamma}(p',q,W)\tau_{\gamma,\delta}(W)
X_{\delta,\beta}(q,p,W) \label{agsx}
\end{eqnarray}
Here $\alpha,\beta$ are specified by the Fock-space of 
the three particles
and the quantum number of the interacting pair. 
The coefficients $C^{1,2}_{\alpha,\beta}$ 
are the spin-isospin recoupling
coefficients given in Ref.~\cite{ikeda}.

The energy of the strange dibaryon resonance
is determined by searching for a pole 
in the scattering amplitude $X$.
To this end, the amplitude is analytically continued
to the unphysical sheet by choosing 
an appropriate path of momentum integration,
and then a pole in the amplitude is located
using the eigenvalue of the kernel $Z\tau$ 
in the above equation;
see Ref.~\cite{ikeda,Glock,Matsu,Pear,Afnan,OT}.

\subsection{Approximate treatment of three-body dynamics}


In the AGS equation,  the three-particle  dynamics is 
incorporated in the particle exchange mechanism $Z$ and 
the  propagator $\tau$. 
The former is 
the three-body interaction
and the latter is determined by two-body t-matrix 
in the presence 
of a spectator particle. 
We first examine how the two-body t-matrix
in a three-body system differs from that in a free-space.
The  t-matrix of $\bar{K}N-\pi\Sigma$ scattering
in the three-particle system is described by the following 
$KNN-\pi Y \Sigma$ coupled-channels equations
\begin{eqnarray}
t_{\alpha,\beta}(W)&=& v_{\alpha,\beta} + \sum_\gamma v_{\alpha,\gamma}
G_0^{\gamma N}(W)
t_{\gamma,\beta}(W), \label{tkn-2}
\end{eqnarray}
where $\alpha,\beta, \gamma = \bar{K}N$ and $\pi\Sigma$, and
the Green function is 
\begin{eqnarray}
G_0^{\alpha N}(W) &=& \frac{1}{W - E_N(\vec{p}_N) -
 \sqrt{(E_{M_\alpha}(\vec{q})+E_{B_\alpha}(\vec{q}))^2 
+ \vec{p}_N^{\ 2}} + i \epsilon}.
\label{green3}
\end{eqnarray}
Here $\vec{p}_N$ is the momentum of the spectator 
nucleon and $\vec{q}$ is
the relative momentum of meson ($M_\alpha$) and baryon ($B_\alpha$) in
the center of mass system of the channel $\alpha$.
The spectator momentum shifts the effective
scattering energy from $W - m_N$ to $W - E_N(p)$ and 
modifies  the `on-shell' momentum of 
the $\pi\Sigma N$ scattering state.
One therefore expects that the motion of the spectator plays 
an important role 
in calculating the binding energy and width of the resonance.

The `isobar' propagator $\tau_{\alpha,\beta}(W)$ 
in Eq. (\ref{agsx})
is  related to  the above t-matrix as
\begin{eqnarray}
<\vec{q}_\alpha|t_{\alpha,\beta}(W)|\vec{q}_\beta>
& = &
g_\alpha(\vec{q}_\alpha)\tau_{\alpha,\beta}(W)g_\beta(\vec{q}_\beta).
\end{eqnarray}
Note that $\tau_{\alpha,\beta}(W)$ depends
on the momentum of the spectator nucleon
through  the three-body Green function in Eq. (\ref{green3}).
Clearly, the effects of the spectator motion 
on $\tau_{\alpha,\beta}(W)$
depend on the momentum distribution of the spectator nucleon,
which can be determined only by solving three-body dynamics.

As mentioned, in `EPA' 
the three-body problem
is treated within the $\bar{K}NN$ Fock-space.
To make contact with `EPA', 
we rewrite Eq. (\ref{tkn-2})
by eliminating $\pi\Sigma N$ state,
which results in the introduction
of the effective interaction $v_{eff}$. 
Thus
\begin{eqnarray}
t_{\bar{K}N-\bar{K}N}(W) &=& 
v_{eff}(W,\vec{p}_N)
+v_{eff}(W,\vec{p}_N) G_0^{\bar{K}NN}(W) t_{\bar{K}N-\bar{K}N}(W).
\label{eff-t}
\end{eqnarray}
The effective interaction $v_{eff}$ is defined by
\begin{eqnarray}
v_{eff}(W,\vec{p}_N) &=& v_{\bar{K}N-\bar{K}N} +
v_{\bar{K}N-\pi\Sigma}
G_0^{\pi\Sigma N}(W)
(1+\bar{t}_{\pi\Sigma-\pi\Sigma}(W) G_0^{\pi\Sigma N}(W))
v_{\pi\Sigma-\bar{K}N},
\label{eff-v}  \\
\bar{t}_{\pi\Sigma-\pi\Sigma}(W) &=& 
v_{\pi\Sigma-\pi\Sigma}
+v_{\pi\Sigma-\pi\Sigma} G_0^{\pi\Sigma N}(W) \bar{t}_{\pi\Sigma-\pi\Sigma}(W).
\label{eff-v2} 
\end{eqnarray}
Note that $\bar{t}$ involves only rescattering through 
the $\pi\Sigma$ interaction.
Solving the above set of equations 
is still equivalent to solving the original Faddeev equation.
The difficulty of treating three-body continuum ($\pi \Sigma N$)
is hidden in the effective potential $v_{eff}$,
which is a three-body interaction
that depends on 
the momentum of the spectator nucleon through 
the Green function.

A drastic simplification of $v_{eff}$
can be achieved 
by neglecting the momentum dependence of the spectator,
or more explicitly, by approximating 
the $\pi\Sigma N$ Green function,  
$G_0^{\pi\Sigma N}(W)$,
in Eqs. (\ref{eff-v}) and (\ref{eff-v2}) with
\begin{eqnarray}
G_{0, approx}^{\pi\Sigma N}(W) &=& \frac{1}{W - m_N -
   E_\pi(\vec{q}) - E_\Sigma(\vec{q}) + i \epsilon}.
\label{green3ap}
\end{eqnarray}
This approximate treatment of the three-body dynamics
represents `EPA' as derived from the Faddeev formalism.

\section{Model of $\bar{K}N$ interaction}

We use here the models
developed in our previous work\cite{ikeda} 
for describing the $\pi N$, $\bar{K}N-\pi\Sigma-\pi \Lambda$
and $NN$  interactions.
Here we briefly explain the
 $\bar{K}N-\pi\Sigma$ interaction in the $I=0$ s-wave channel,
which plays a crucial role in our study of  the strange dibaryons.
 Our starting point is the following 
leading order effective chiral Lagrangian
for a baryon $\psi_B$ and a pseudoscalar meson $\phi$,
\begin{eqnarray}
L_{int} & = & \frac{i}{8F_\pi^2}tr(\bar{\psi}_B \gamma^\mu
                                 [[\phi,\partial_\mu \phi],\psi_B]).
\end{eqnarray}
The s-wave meson-baryon potential derived from $L_{int}$ is of
the following separable form
\begin{eqnarray}
<\vec{p}',\beta|V_{BM}|\vec{p},\alpha>
 & = & - C_{\beta,\alpha}\frac{1}{(2\pi)^3 8F_\pi^2}
\frac{m_\beta + m_\alpha}
{\sqrt{4E_{\beta}(\vec{p}^{\ \prime})E_\alpha(\vec{p})}} \nonumber \\ & \times &
g_\beta(\vec{p}^{\ \prime})g_\alpha(\vec{p}),
\label{mb-int}
\end{eqnarray}
with $g_\alpha(\vec{p})=\Lambda_{\alpha }^4/(\vec{p}^{\ 2} + \Lambda_{\alpha }^2)^2$.
Here $\vec{p}$ and $\vec{p}^{\ \prime}$ are the momenta of the mesons in the
initial $\alpha$ and  the final  $\beta$ states, respectively.
The coupling constants $C_{\beta,\alpha}$ are :
 $C_{\bar{K}N-\bar{K}N}=6,C_{\bar{K}N-\pi\Sigma}=-\sqrt{6}$
and $C_{\pi\Sigma-\pi\Sigma}=8$.
Thus the potential is attractive for both $\bar{K}N$ and $\pi\Sigma$ channels.
The strength of the potential at zero momentum is 
determined by the pion decay constant $F_\pi$($F_\pi = 92.4MeV$).
We found  only one resonance  in 
the $I=0$ $\bar{K}N-\pi\Sigma$ s-wave channel.
Meanwhile it has been shown in Refs. \cite{HW,Jido,CG07} that
two resonance poles may exist in this channel.
The potential in Eq. (\ref{mb-int}) is independent of 
the scattering energy. The energy dependence of the $\bar{K}N-\pi\Sigma$
interactions is one of the important features that differentiate our approach
from the chiral unitary approach. The properties of  the resonances
and the energy dependence of the potentials 
in the s-wave meson-baryon scattering will be examined 
in Ref.~\cite{ikeda3}.

\begin{figure*}
\resizebox{0.5\textwidth}{!}{%
\includegraphics{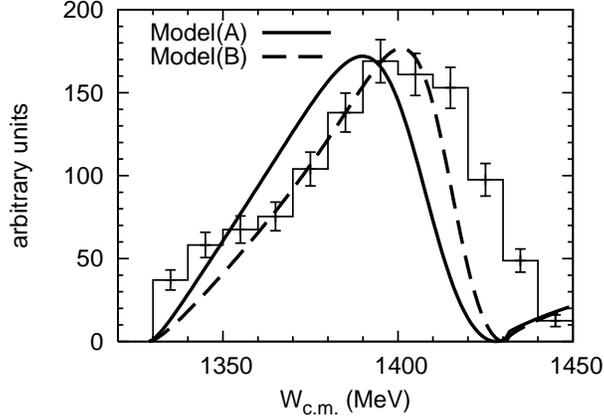}}
\caption{Invariant mass distribution of the $\pi \Sigma$.}
\label{inv-ps}
\end{figure*}

\begin{figure*}
\resizebox{1.0\textwidth}{!}{%
\includegraphics{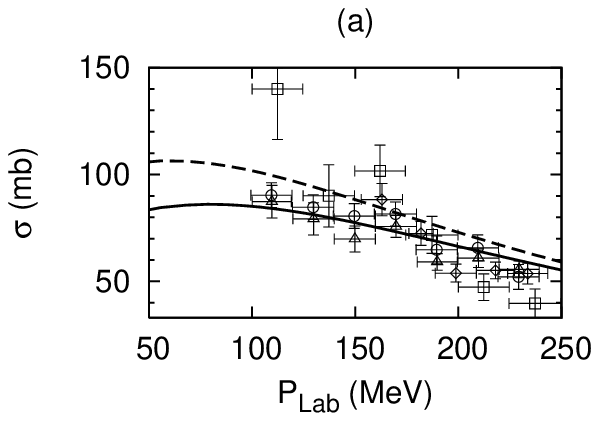}
\includegraphics{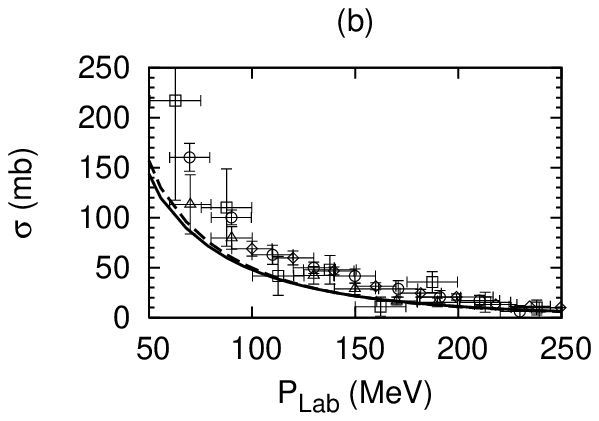}
\includegraphics{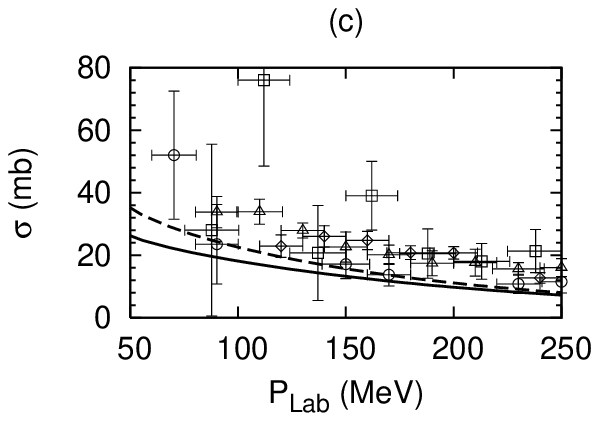}
}\caption{The total cross section of 
(a) $K^-p \rightarrow K^-p$,
(b) $K^-p \rightarrow \pi^+\Sigma^-$, and
(c) $K^-p \rightarrow \pi^-\Sigma^+$ reactions.
The solid (dashed) curve shows cross section calculated by using
Model (A)((B)). Data are taken from 
Ref.~\cite{crs1,crs2,crs3,crs4,crs5}.
}
\label{kp-cros}
\end{figure*}

We first discuss the values of the
cutoff parameters, $\Lambda_\alpha$'s, 
to be used in this work.
Table \ref{hyo1} gives two sets
of choices, Model (A) in the first row and 
Model (B) in the second row.
The cutoff parameters for Model (A) are taken from 
Model (f) in Ref.~\cite{ikeda}, which was constructed to generate
a resonance at around $1405$MeV \cite{Dlitz}.
When the couplings between the $\bar{K}N$ and 
the $\pi\Sigma$ are switched off 
in Model (A),
 a bound state appears in the $\bar{K}N$ channel and
there is no resonance in the $\pi\Sigma$ channel.
To test the prediction of Model (A) in the energy region below
the $\bar{K}N$ threshold, 
we study the $\pi^-\Sigma^+$ mass distribution in 
the $K^-p$ reaction~\cite{Hem}.
Following Ref. \cite{veit}, we calculate
the $\pi^-\Sigma^+$ mass distribution 
from  the $I=0$ $\pi \Sigma$ scattering t-matrix
\begin{eqnarray}
\frac{dN}{dW_{c.m.}}=C |t_{\pi\Sigma-\pi\Sigma}|^2 p_{c.m.},
\label{massdist}
\end{eqnarray}
where $p_{c.m.}$ is the $\pi \Sigma$ relative momentum 
in the center of mass system.
Because of the presence of an arbitrary 
constant $C$, only the shape of the
mass distribution can be compared with the data.
We assume that
the mass distribution of  $\pi^-\Sigma^+$ is dominated 
by the $I=0$ amplitude and 
 that the mass distribution can be deduced
from the $\pi\Sigma$ rescattering,
neglecting other energy dependence due to 
the $\pi\Sigma$ production mechanism.
The mass distribution calculated  from the model (A) is compared with
the data in Fig.~\ref{inv-ps}. Model (A) gives a spectrum
slightly larger than the data in the lower mass region.
To examine the model dependence of this analysis,
we study Model (B), which
gives a slightly better description of the $\pi^-\Sigma^+$
mass distribution. 
The results for Model (B) are shown  
in dashed line in Fig. \ref{inv-ps}.
The Model (B) has a slightly weaker $\bar{K}N$
interaction than Model (A) because of 
the smaller value of the cutoff parameter.
The resonance generated from Model (B)
is less bound and has a narrower width than that 
generated from Model (A).
Both models give a satisfactory description
of the total cross sections for the 
$K^-p \rightarrow K^-p$ reaction
[Fig. \ref{kp-cros}(a)],
$K^-p \rightarrow \pi^+\Sigma^-$ reaction
[Fig. \ref{kp-cros}(b)], and
$K^-p \rightarrow \pi^-\Sigma^+$ 
[Fig. \ref{kp-cros}(c)] reaction.

\begin{table*}[htbp]
\begin{center}
\begin{tabular}{c||cc|cc}
   &$\Lambda_{\bar{K}N}$(MeV)&$\Lambda_{\pi\Sigma}$(MeV)& 
Resonance energy(MeV)\\ \hline \hline
(A) & 1160 & 1100 & $1405.8-i25.2$\\ 
\hline
(B) & 1100 & 1100 & $1414.2-i18.6$\\ 
\hline
\end{tabular}
\caption{The cutoff parameters
and the resonance energy of the $\Lambda(1405)$.}
\label{hyo1}
\end{center} 
\end{table*}

\section{Results and Discussion}

The energies of the strange dibaryon resonances obtained
from our Faddeev approach described in section II are
listed in the first row of Table \ref{hyo2}.
The half width of the resonance is about 22 MeV 
(23 MeV)
for Model (A)((B)).  
Model (B) gives a binding energy
($-B = Re(W_{pole}) - m_K - 2m_N$)
about 20MeV smaller than Model (A).

In all of the previous theoretical studies of strange dibaryon
resonances except our previous work\cite{ikeda},
the $\pi \Sigma N$ Fock space is not treated explicitly; 
it is only included in the intermediate states  of 
the two-body $\bar{K}N-\pi\Sigma$ scattering amplitude. 
The influence of this simplification can 
be examined in our approach by 
turning off the pion exchange $Z$ term in Eq. (\ref{Eqags-2}).
As can be seen in Table \ref{hyo2},
the pion-exchange $Z$ term plays
only a minor role 
in the determination of the resonance energy. 
In the following discussion therefore
we will treat the Faddeev calculation
without the pion-exchange $Z$ term
as the `Exact' calculation.

\begin{table*}[htbp]
\begin{center}
\begin{tabular}{c||c|c}
& Model (A)     & Model (B) \\
\hline \hline
Full calculation & $-63.3-i22.2$ & $-44.4-i22.8$ \\
\hline
without pion-exchange $Z$ & $-66.9-i21.7$ & $-47.4-i25.0$ \\
\hline
\end{tabular}
\caption{The pole energy of the strange dibaryon  resonance is given in MeV.
The pole energy is related to the binding energy $B$ and the width $\Gamma$
as $W_{pole} - m_K - 2 m_N = - B-i \Gamma /2$.
}
\label{hyo2}
\end{center} 
\end{table*}

\begin{table*}[htbp]
\begin{center}
\begin{tabular}{c||c|c}
    & Model (A)     & Model (B) \\
\hline \hline
`Exact'  &  $-66.9-i21.7$ & $-47.4-i25.0$ \\
EPA  & $-41.8-i35.4$ & $-31.5-i26.3$ \\
\hline
\end{tabular}
\caption{The pole energies obtained from `EPA' are compared with
the `Exact' results;
for the explanation of the term `Exact' see the text.}
\label{hyo3}
\end{center} 
\end{table*}

\begin{figure*}
\resizebox{0.9\textwidth}{!}{\hspace*{-0.5cm}
\includegraphics{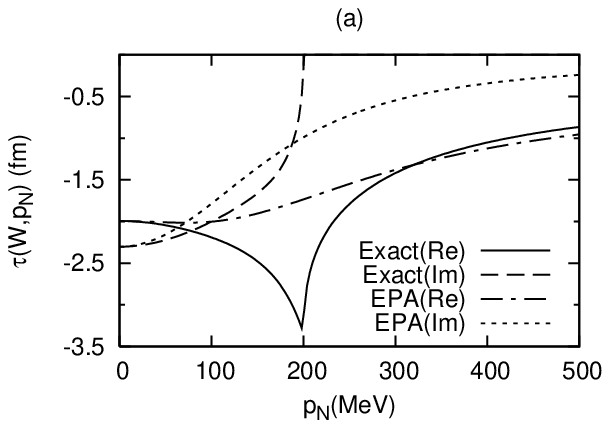}\hspace*{0.5cm}
\includegraphics{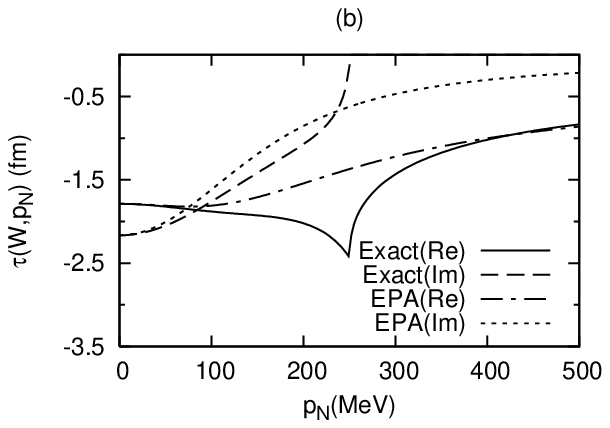}}
\caption{$\bar{K}N$ amplitude of (a) Model (A) and 
(b) Model (B).
 $\bar{K}N$ amplitudes for $I=0$ are shown for `Exact' and
`EPA'
treatments of the $\pi\Sigma N$ Green function.
}
\label{off-tau}
\end{figure*}

Table \ref{hyo3} provides comparisons 
between the resonance energy obtained
in the `Exact' calculation
and that obtained
from `EPA' described in section II.
Clearly,
there are significant differences
between the two approaches.
The `EPA' calculation gives 
a binding energy 15 to 25 MeV smaller
than the `Exact' calculation.
Similar effects of the three-body dynamics
were partly studied in  Ref. \cite{WG}.
To understand these results, 
it is informative to plot $\tau(W)$ defined in Eq. (\ref{tau})
as a function of the momentum $p_N$ of the spectator nucleon 
in the most important $I=0$ $\bar{K}N$ channel;
see Fig.~\ref{off-tau}.
The amplitude $\tau(W)$ is evaluated at
$B=66.9 (47.4)$ MeV for Model (A)((B)).
The real and imaginary  parts of $\tau$ are
shown in solid (dash-dotted) and dashed (dotted)
curves for the `Exact' (`EPA') calculations.
As $p_N$ increases,
the scattering energy available for 
the $\pi\Sigma$ system decreases.
This implies that in the `Exact' calculation
the effects of the $\pi\Sigma$ threshold
appear as a cusp in the real part of $\tau$ at `threshold'
and the vanishing of the imaginary part of $\tau$  for 
the larger value of $p_N$, see Fig. \ref{off-tau}.
On the other hand, `EPA' fails to capture
this important behavior of the t-matrix.

As mentioned,
`EPA' involves the approximation of the $\pi\Sigma N$ Green function.
If we further approximate the $\bar{K}NN$ Green function
we are led to the $t\rho$ approximation, 
which underlies the first order optical potential model.
With this additional approximation
we find a resonance at $(-67.4-i64.2)$ MeV
and $(-60.6 - i47.7)$ MeV for Model (A) and (B), respectively.
We note that the additional approximation
influences the resonance width drastically.

In summary, our analysis 
clearly shows that the exact treatment
of three-body dynamics, such as given by the Faddeev formulation, 
is essential in making precise predictions 
on the resonance positions
of the strange dibaryons.

Finally, we examine the behavior of the resonance pole  
trajectory as the magnitude of $\bar{K}N$ interaction
is artificially increased from its physical value.
Let $f$ stand for an enhancement factor of 
strength of the $I=0$ $\bar{K}N$ interaction: 
\begin{eqnarray}
\bar{v}_{\bar{K}N,\bar{K}N} 
& = & f v_{\bar{K}N,\bar{K}N}. \label{factf}
\end{eqnarray}
The resonance determined 
from the pole of the scattering amplitude
in our $\bar{K}N-\pi\Sigma$ coupled-channels model becomes
a `virtual state' as $f$ increases.
This behavior of the two-body resonance pole is similar to the one
observed in Ref.~\cite{Aka3}. 
Although it was discussed in Ref. \cite{Aka3}
that the spectrum shape of the Green function
cannot be well explained based on the pole 
of the Green function, we emphasize that
the spectrum shape can in fact be well described
in terms of the resonance pole in the amplitude,
we take into account the residue at the pole 
and the next order term in the
Laurent expansion of the Green function; 
see Ref. \cite{suzuki}.

\begin{figure*}
\resizebox{0.8\textwidth}{!}{%
\includegraphics{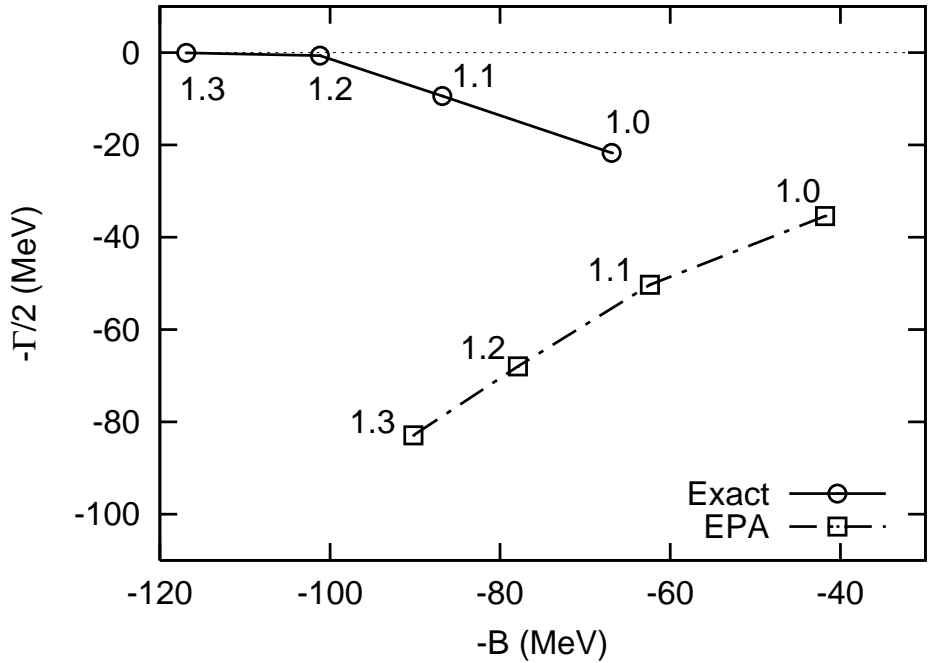}}
\caption{Resonance energy of $\bar{K}NN-\pi Y N$ system for Model (A).
The circles (squares)
show resonance energies obtained in
the `Exact' (`EPA')
treatment of the three-body Green function.
The numbers attached to the circles and squares 
give the corresponding values of
the enhancement factor $f$ in Eq.~(\ref{factf}). 
}
\label{mod-kn}
\end{figure*}

The trajectory of the  resonance pole 
occurring in the three-body system
behaves quite differently from the resonance pole 
in the two-body system.
The resonance energies obtained 
from our Model (A) 
are shown as circles in Fig. \ref{mod-kn}. The squares
in the same figure correspond to the `EPA' results.
The numbers attached to the circles and squares 
give the corresponding values of the enhancement factor $f$.
As $f$ increases,
the binding energy of the resonance increases
for both the `Exact' and `EPA' cases.
In the `Exact' calculation (circles),
the imaginary part of the resonance energy becomes smaller 
as the binding energy increases and, for $f=1.3$,  
the resonance almost becomes a bound state.
On the other hand, in the `EPA' case (squares)
the resonance becomes a virtual state as $f$ grows.
By contrast, in the 'Exact' case
the resonance energy of the three-body system
determined from the pole of the
scattering amplitude does not become a virtual state 
even for an (artificially) strong strong $\bar{K}N$ 
interaction.
Here again we see the importance of 
taking a full account of three-body dynamics
for understanding the strange dibaryon resonances.


\section{Summary}

We have demonstrated the importance of 
taking a proper account of three-body dynamics
in predicting the resonance energies of strange dibaryons.
Within the Faddeev formulation 
we have examined the approximations
involved in the existing effective potential approach
(which for short we refer to as `EPA').
Upon eliminating the $\pi \Sigma N$ Fock space,
the effective interaction in the $\bar{K}NN$ sub-space 
becomes a three-body interaction which depends 
on the resonance energy and the momenta 
of all the three particles. 
We have shown that this energy and momentum dependence
(which is neglected in `EPA')
plays a important role in determining the resonance energies.
As regards the behavior of the resonance position
as a function of the strength
of the $\bar{K}N$ potential, 
we have shown that the appearance of a virtual state
in `EPA' as the strength of the $\bar{K}N$ potential
grows
is an artifact of the approximations involved in `EPA'.
We have demonstrated that the results
obtained from the Faddeev approach indicate that
the resonance becomes a bound state as 
the $\bar{K}N$ potential becomes strong.
In conclusion, we emphasize that a full treatment of 
three-body dynamics is
essential in understanding 
the $\bar{K}NN-\pi Y N$ coupled-channels resonance.

\begin{acknowledgments}

The authors would like to thank Drs. T.-S. H. Lee and
N. Suzuki for critical discussions on the resonance pole.
The author's thanks are due to Prof. K. Kubodera
for reading the manuscript and making useful suggestions.
This work is supported by the Japan Society for the Promotion of Science,
Grant-in-Aid for Scientific Research(c) 20540270.

\end{acknowledgments}


\end{document}